\documentclass[journal,onecolumn,12pt]{IEEEtran}

\ifCLASSINFOpdf
\else
   \usepackage[dvips]{graphicx}
\fi
\usepackage{url}

\usepackage{setspace}

\usepackage{soul}
\usepackage{color, xcolor}

\usepackage{setspace}

\usepackage{graphicx}

\usepackage{subfigure} 

\usepackage{amssymb}

\usepackage{amsmath}

\usepackage{amsfonts}

\usepackage{graphicx}

\usepackage{epstopdf}

\usepackage{float}

\usepackage{stfloats}

\usepackage{indentfirst}

\usepackage[explicit]{titlesec}

\usepackage[citecolor=red]{hyperref}

\titlespacing*{\section}{0pt}{0.5ex plus .0ex minus .0ex}{0.5ex plus .0ex}
\titlespacing*{\subsection}{0pt}{0.5ex plus .0ex minus .0ex}{0.5ex plus .0ex}
\begin{document}
\title{Distributed Neural Precoding for Hybrid mmWave MIMO Communications with Limited Feedback}

\author{Kai Wei,
Jindan Xu,
Wei Xu, \IEEEmembership{Senior Member, IEEE,}
Ning Wang,
and Dong Chen
\thanks{K. Wei, J. Xu, and W. Xu are with the National Mobile Communications Research Laboratory (NCRL), Southeast University, Nanjing 210096, China. W. Xu is also with Henan Joint International Research Laboratory of Intelligent Networking and Data Analysis, Zhengzhou University, Zhengzhou, 450001 China (e-mail: \{kaiwei, jdxu, wxu\}@seu.edu.cn).}
\thanks{N. Wang is with the Department of Electrical Engineering, Zhengzhou
University, Zhengzhou, Henan 450001, China (e-mail: ienwang@zzu.edu.cn).
}
\thanks{D. Chen is with Xiaomi Corporation, Beijing 100085, China (chendong7@xiaomi.com).}}

\maketitle
\vspace{-1.6cm}
\begin{abstract}
Hybrid precoding is a cost-efficient technique for millimeter wave (mmWave) massive multiple-input multiple-output (MIMO) communications. This paper proposes a deep learning approach by using a distributed neural network for hybrid analog-and-digital precoding design with limited feedback. The proposed distributed neural precoding network, called \emph{DNet}, is committed to achieving two objectives. First, the DNet realizes channel state information (CSI) compression with a distributed architecture of neural networks, which enables practical deployment on multiple users. Specifically, this neural network is composed of multiple independent sub-networks with the same structure and parameters, which reduces both the number of training parameters and network complexity. Secondly, DNet learns the calculation of hybrid precoding from reconstructed CSI from limited feedback. Different from existing black-box neural network design, the DNet is specifically designed according to the data form of the matrix calculation of hybrid precoding. Simulation results show that the proposed DNet significantly improves the performance up to nearly 50\% compared to traditional limited feedback precoding methods under the tests with various CSI compression ratios.
\end{abstract}

\IEEEpeerreviewmaketitle

\section{Introduction}

Millimeter wave (mmWave) massive multiple-input multiple-output (MIMO)  communication systems utilize the wide bandwidth of mmWave to achieve high transmission rates. It has been a key technology in the fifth-generation mobile communication system \cite{bib1}. One of the major challenges faced by mmWave massive MIMO is that, with the rapid increase in the number of antennas, equipping each antenna with a separate radio-frequency (RF) chain can result in high power consumption and hardware cost. This challenge makes the traditional fully-digital precoding architecture difficult to implement in practice. To address this issue, hybrid analog-and-digital precoding was proposed for mmWave MIMO system \cite{bib2}-\cite{bib5}. In the hybrid precoding architecture, expensive fully-digital precoder is replaced by a network of low-cost phase shifters, called analog precoder, followed by a low-dimensional digital precoder.

However, the design of hybrid precoding is a challenge even when the base station (BS) obtains perfect channel state information (CSI). Nonconvex constant modulus constraints are imposed on the analog precoding parameters, which come from the physical constraints of the phase shifter network. Due to this nonconvexity, the optimization of hybrid precoding is difficult to solve by traditional convex optimization tools.
Iterative algorithms, e.g., \cite{bib2}-\cite{bibnew1} by using alternating optimization were proposed for calculating the hybrid precoding, which aimed at achieving near-optimal performance in the case of narrowband or wideband. 
Alternatively, researchers tried to solve the hybrid precoding design by exploiting deep learning (DL) technologies \cite{bib8}-\cite{bib11}. 
In \cite{bib10}, the authors transformed the channel matrix to an image and then designed convolutional neural networks (CNN) for the precoding design in massive MIMO. 
The study in \cite{bib11} introduced an unsupervised DL-based approach to calculate the hybrid precoding and achieved performance improvement compared to the supervised DL-based methods.

The above methods all require the BS to know perfect CSI. 
However, in a frequency division duplexed (FDD) system, CSI is estimated by receivers and then fed back to the BS via a limited feedback link.
As the number of antennas increases, the channel matrix dimensions and the number parameters to feedback rise sharply.
Therefore, it is of practical interest to design hybrid precoding with limited feedback. 
The authors of \cite{bib6} proposed an analog precoding design from a given CSI codebook so that only a low-dimensional equivalent matrix in the form of a codeword is fed back. An alternative solution is to apply compressed sensing (CS) \cite{bib12} or DL-based data compression methods \cite{bib13}-\cite{bib15} to compress and reconstruct the CSI. Deep reinforcement learning (DRL) based beam pattern design was presented in the \cite{bibnew2}, which learned the beam pattern without the explicit knowledge of the channels.
The authors of \cite{bibnew3} combined channel estimation with hybrid precoding design and proposed a quantized network to calculate quantized analog precoding.

In this study, we propose a distributed neural network, called \emph{DNet}, for the hybrid precoding design with reduced number of channel feedback parameters. 
The design of DNet is based on a distributed architecture of neural networks, which can be distributively deployed at the users and BS after the network training is completed to achieve implicit CSI compression and feedback. 
The network that performs the CSI compression is composed of multiple sub-networks with the same structure and parameters, so that the number of training parameters of the entire neural network, as well as the network complexity, are noticeably reduced. 
Moreover, the sub-network responsible for hybrid precoding calculation is different from traditional black-box-based DL methods \cite{bib10}.
By exploiting the characteristics of the matrix in hybrid precoding, we have structured the design of this sub-network to achieve improved performance.

The rest of this paper is organized as follows. In Section \uppercase\expandafter{\romannumeral2}, we describe the system model and problem formulation. The architecture of the proposed DNet is introduced in Section \uppercase\expandafter{\romannumeral3}. Section \uppercase\expandafter{\romannumeral4} presents the simulation results and Section \uppercase\expandafter{\romannumeral5} concludes this paper.

\section{System Model and Problem Definition}

We consider the downlink of a multiuser massive MIMO system as shown in Fig. 1. The BS is equipped with $N_t$ antennas and $N_{RF}$ RF chains, serving $K$ single-antenna user-equipments (UE) where $K \le N_{RF}$. We assume the BS serves the most users, i.e., $N_{RF} = K$.

In the downlink, the hybrid precoding contains a digital baseband precoder and an analog precoder, denoted by ${\bf D}$ of  dimension $K \!\times\! K$ and $\bf F$ of dimension $N_t \!\times\! K$, respectively. 
The digital baseband precoder, $\bf D$, adjusts both the amplitude and phase of the input signals. The analog precoder $\bf F$, which is realized by variable phase shifters, adjusts only the phase of the input signals.
This imposes the constraint that the modulus of every element in $\bf F$ is a constant, i.e., $|[{\bf F}]_{i,j}|=\frac{1}{\sqrt{N_t}}$, where $[{\bf F}]_{i,j}$ is the $(i,j)$-th element of $\bf F$.

Assuming a flat fading channel, the signal received by the $k$th user can be expressed as
\begin{equation}
	y_k={\bf h}_k^H{\bf{F}}{\bf d}_k{s}_k+\sum_{j\not=k}{\bf h}^H_k{\bf{F}}{\bf d}_j{s}_j+n_k,
\end{equation}
where ${\bf h}^H_k$ denotes the channel from the BS to the $k$th user, ${\bf{d}}_j$ is the $j$-th column of  digital precoding $\bf D$, and $n_k$ represents the additive white Gaussian noise (AWGN) with unit variance. ${\bf s} = [s_1, ..., s_k]^T\in \mathbb{C}^{K \times 1}$ represents the signal vector for all $K$ users, subjecting to the power limit, i.e., $\mathbb{E}[{\bf s}{\bf s}^H]=\frac{P}{K}{\bf I}_K$, where $P$ is the BS transmit power and $\mathbb{E}[\cdot]$ is the expectation operator. Considering the transmit power constraint, we normalize the energy of $\bf F$ and $\bf D$ to satisfy $\|{\bf FD}\|_{\rm F}^2=K$. Then, the sum rate of the system is calculated by 
\begin{equation}
	R=\sum_{k=1}^{K}\log_2 (1+{\rm SINR}_k),
\end{equation}
where ${\rm SINR}_k$ implies the signal-to-interference-plus-noise-ratio (SINR) of $k$th user and is follows
\begin{equation}
	{\rm SINR}_k=\frac{\frac{P}{K}|{\bf h}^H_k{\bf F}{\bf d}_k|^2}{1+\sum_{j\not=k}\frac{P}{K}|{\bf h}^H_k{\bf F}{\bf d}_j|^2},
\end{equation}
where the ${\bf d}_i$ for $i=1,...,K$ is the $i$th column of $\bf D$.
Now the optimization problem of the hybrid precoding in the multiuser massive MIMO system is formulated as
\begin{equation}
	\begin{aligned}
\{ {\bf F}, {\bf D} \} &= \arg\,\mathop{\max}_{\bf F,\bf D}\,R,  \\
&\begin{array}
{r@{\quad}r@{}l@{\quad}l} \rm{s.t.}
&\qquad\;|{\bf F}_{i,j}|=\frac{1}{\sqrt{N_t}},\\
&\,\|{\bf FD}\|_{\rm F}^2=K, \\
\end{array} 
\end{aligned}
\end{equation}
where $|.|$ and $\|.\|_{\rm F}$ respectively represent the modulus and Frobenius norm operations.
In an FDD system, the BS calculates the precoder matrices, $\bf F$ and $\bf D$, by using algorithms like in \cite{bib2}-\cite{bib7} to solve the problem in (4), where perfect CSI of $\bf H$ is available at the BS. 
However, there are a large number of antennas in a massive MIMO system, and the total number of the feedback parameters grows like $N = 2KN_t$, which implies that the number of feedback parameters would be too large for a feedback link with limited bandwidth.

To reduce the number of feedback CSI parameters, it is necessary to compress the channel matrix by an encoder in the UE and try to accurately decode the channel matrix at the BS. 
The encoder at the UE is denoted by 
\begin{equation}
	 {{\bf {q}}_k} = f_{\rm EN}({\bf{h}}_k), 
\end{equation}
where ${\bf q}_k$ is a $2M_t$-dimensional compressed codeword of ${\bf{h}}_k$ and $f_{\rm EN}(\cdot)$ is the compression operation of the encoder,
which reduces the number of feedback parameters from $N$ to a much smaller value of $M = 2KM_t$. The compression ratio is defined as $\gamma = M/N $. Similarly we recover the channel matrix at the BS through a decoder $f_{\rm DE}(\cdot)$, 
\begin{equation}
	 {\widetilde{\bf {h}}_k} = f_{\rm DE}({\bf q}_k). 
\end{equation}
According to the above procedure of CSI compression and feedback, the channel matrix retrieved at the BS is
\begin{equation}
	{\widetilde{\bf {h}}_k} = f_{\rm DE}(f_{\rm EN}({\bf h}_k)) = {\bf h}_k + \widetilde{\bf z},
\end{equation}
where $\widetilde{\bf z}$ represents the CSI error due to the compression and reconstruction. 
In fact, the BS designs the hybrid precoding based on imperfect CSI, rather than perfect CSI. 

To solve the problem in (4) with the imperfect CSI, we propose a method by using a distributed neural network, named DNet, to reduces the number of feedback parameters from the UEs, and design the hybrid precoder implicitly using an imperfect version of the CSI at BS. 
\begin{figure*}[tb]
\centering
\includegraphics[width = .7\textwidth]{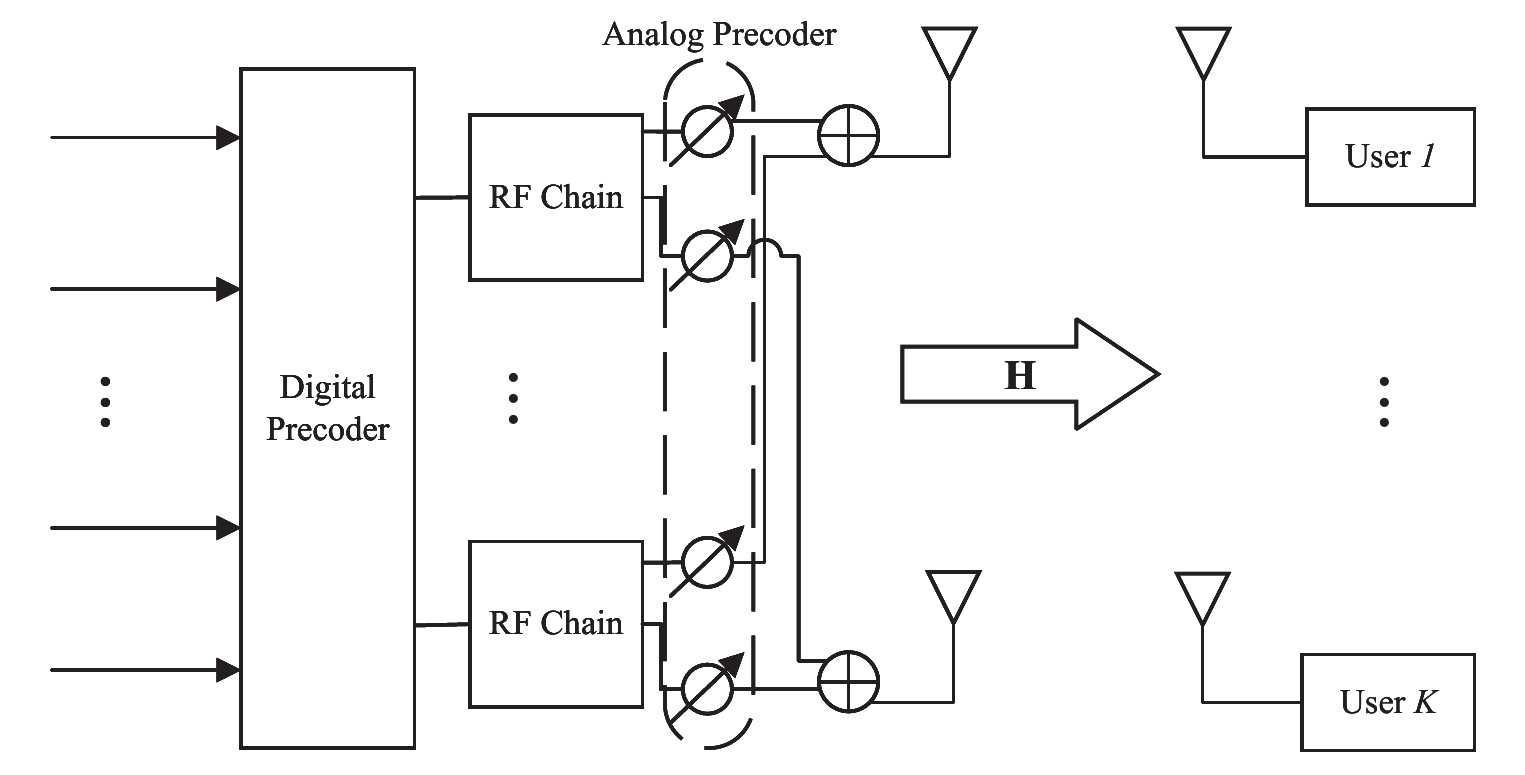}
\caption{System model of the hybrid mmWave precoding.}
\label{fig:myphoto}

\end{figure*}

\section{Design of DNet}

\begin{figure}[H]
\centering
\includegraphics[width = 0.9\textwidth]{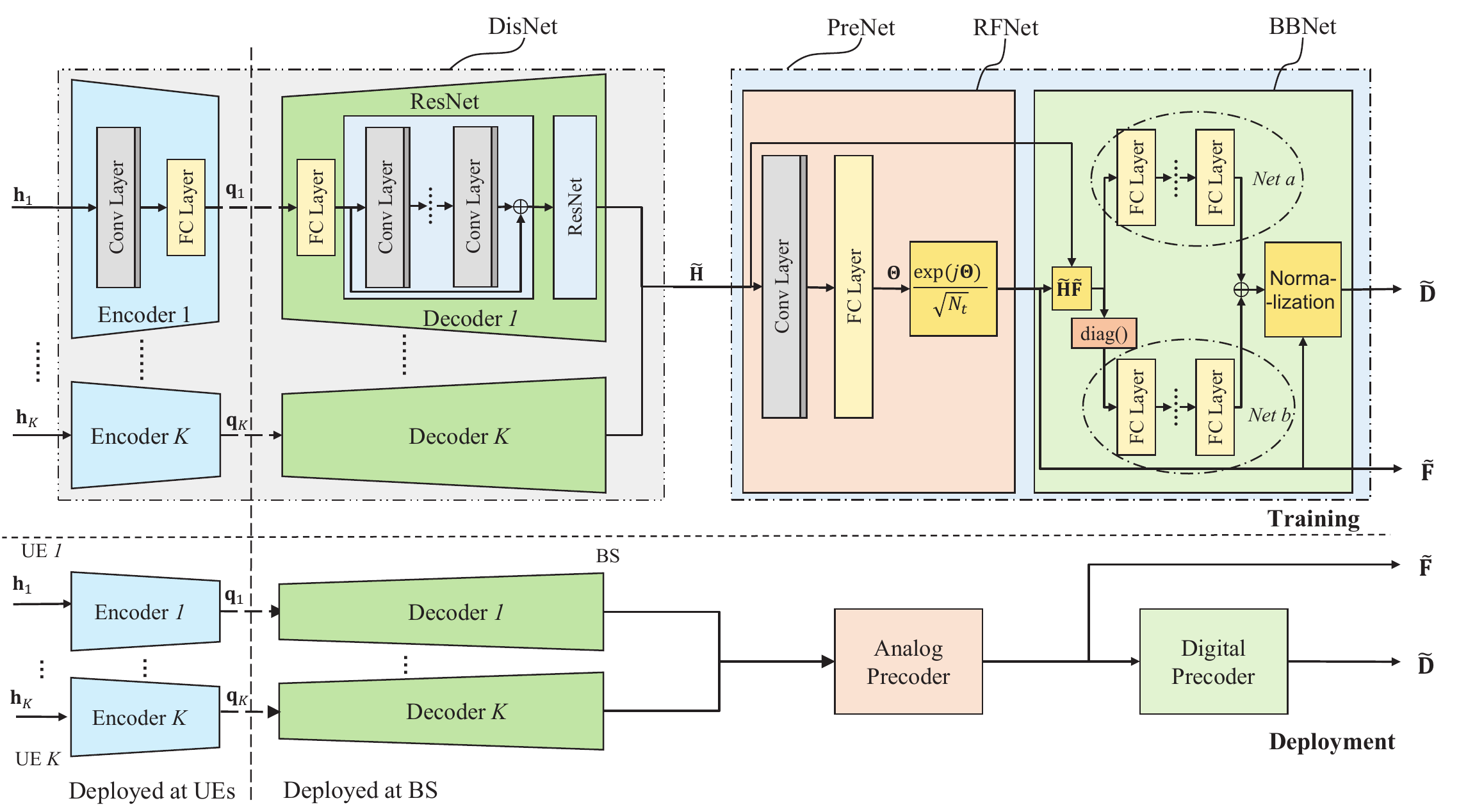}
\caption{Proposed architecture of the DNet.}
\label{fig:myphoto}
\end{figure}

In this section, we elaborate the new architecture of the proposed DNet to calculate the hybrid precoding for the mmWave MIMO with limited CSI feedback. 
The DNet consists of two subnetworks, which are referred to as DisNet and PreNet. These two subnetwork are respectively responsible for the implicit CSI compression reconstruction and the hybrid precoding design.

\subsection{DisNet for Limited Feedback}
The architecture of the DNet is elaborated in Fig. 2. 
The first component structure of DNet is DisNet, which is split and deployed on the UEs and BS after training. The DisNet consists of $K$ pairs of encoder-decoder network. 
We stress that each of the encoder-decoder network pair shares the same network architecture and parameters. Unlike existing neural network based techniques for CSI compression, e.g., in \cite{bib13}, \cite{bib14}, the proposed DisNet is based on a distributed neural network architecture which is therefore suitable for deployment in multiuser mmWave MIMO. 
In the DisNet, each encoder-decoder network pair works for one UE. The channel matrix between the $k$-th UE and BS, denoted by ${\bf h}^H_k$, is a $1 \!\times\! N_t$-dimensional complex vector. 
Thus, in our proposed encoder-decoder network, we choose a one-dimensional (1D) convolutional (Conv) layer and a fully-connected (FC) layer as the basic unit to compress and recover the CSI. And the activation function is chosen as
\begin{equation}
	{\sigma}{\rm(}x{\rm)} ={\rm LeakyReLU}{\rm(}x{\rm)} = 
\left\{
\begin{array}{rcl}
	x,  & & {x \geq 0} \\
	0.3x, & &{x\; \textless\; 0,}\\
\end{array}
\right.
\end{equation}
to provide non-linearity and avoid dead net problem \cite{bibnew7}.

Concretely for the $k$th encoder-decoder network pair, the encoder consists of one 1D Conv layer and one FC layer. 
The former performs CSI feature extraction, while the latter performs CSI feature data compression.
Thus, ${\bf h}_k$ would be reshaped to $[\mathcal{R}({\bf h}_k),\mathcal{I}({\bf h}_k)]^H$, which is a $2\times N_t$ tensor, where $\mathcal{R}({\bf h}_k)$  and $\mathcal{I}({\bf h}_k)$ are the real and imaginary parts of ${\bf h}_k$, as the inputs of the first Conv layer.
The convolutional kernels in the first Conv layer are of the number $4$ and of the size $ 2 \!\times\! 3$, which is used to extract features of the channel, ${\bf h}_k$, and generates a $ 4\times N_t$ tensor as the output. The output of the Conv layer is then flattened into a $4N_t$-dimensional vector before fed to the FC layer. This layer compresses the CSI feature data and generates a feedback vector ${\bf q}_k$, which is a $2M_t$-dimensional vector, for limited feedback. Thus, the compression ratio can be also calculate as $\gamma = M_t/N_t$. The $K$ encoder parts are trained for once and then distributively deployed at the corresponding UE. Each UE employs its encoder sub-network to generate limited feedback parameters, and sends them back to the BS.

Once the BS receives those codewords, i.e., $\{ {\bf q}_1,\cdots,{\bf q}_K \}$, through limited feedback links from the UEs, we propose a subnetwork of decoder to recover the CSI vectors for all users. 
To account for feedback or quantization errors, we superimpose a noise on the input to the decoder. 
The first layer of the decoder is an FC layer, which converts a $2M_t$-dimensional vector to a $2N_t$-dimensional vector. Then this vector is reshaped to a $2\!\times\! N_t $ tensor, setting as the input of following neural network. 
The following layers include two residual networks, named ResNet. Each of the ResNet consists of four Conv layers. 
From the first Conv layer to the fourth Conv layer, the number of convolutional kernels are respectively set to 4, 8, 16, and 2. The corresponding dimensions of these convolutional kernels are, respectively, chosen as $2\!\times\! 3$, $4\!\times\! 3$, $8\!\times\! 3$, and $16\!\times\! 3$.
We use $\bf x$ and $\bf y$ to respectively represent the input and output of the ResNet while let $\rm{Res}(\cdot)$ denote the operations of the four Conv layers in the ResNet. Then the output of each ResNet can be expressed as
\begin{equation}
	{\bf y}={\bf x}+{\rm Res}({\bf x}).
\end{equation}
Then, all the outputs of the $K$ decoders are combined at the BS to yield the $\widetilde{\bf H}=[\widetilde{{\bf h}}_1^{ H},\widetilde{{\bf h}}_2^{ H},\cdots,\widetilde{{\bf h}}_K^{ H} ]^{ H}$, as depicted in Fig.2, where $\widetilde{\bf h}_k^H$ regarded as the the implicit version of CSI of the $k$-th UE.

Note that the proposed encoder-decoder network pairs are set to the same structure as described above, in terms of not only the architecture but also the value of parameters. 
This same setting for each sub-network helps reduce the complexity of the proposed DNet.
In particular, it is found that the DisNet component of the DNet actually learns the distribution of the channel data set, which allows the DisNet to be extendable.

\subsection{PreNet for Hybrid Precode}
The second component network at the BS is PreNet. As shown in the right part of  Fig. 2, it learns to calculate the precoder according to the implicit CSI, $\widetilde{\bf{H}}$, obtained by the DisNet. Specifically, the PreNet consists of two parts, i.e., RFNet and BBNet, generating analog-precoding and digital-precoding, respectively.

Consider that the input of RFNet, i.e., the implicit CSI, $\widetilde{\bf{H}}$, is a $K\!\times\!{N_t}$-dimensional complex matrix. We then choose a two-dimensional (2D) Conv layer as the first layer in RFNet to extract internal features of $\widetilde{\bf H}$. 
The real and imaginary parts of the $\widetilde{\bf H}$, i.e., $[\mathcal{R}({\bf H});\mathcal{I}({\bf H})]$, are marked with different colors in the image and are input of the 2D Conv layer. The two convolutional kernels of this Conv layer are with the size of $3\!\times\! 3$. And we apply zero padding to make sure that the size of the output feature map remains to be $K\!\times\! N_t$. 
The output is flattened to a vector of size $2KN_t \!\times\! 1$ and then sent to the FC layer. The output of this FC layer is a $KN_t$-dimensional vector and is then reversely reshaped into a $K\!\times\! N_t$-dimensional matrix $\bf\Theta$. Then the analog precoding matrix from RFNet is expressed as
\begin{equation}
	\vspace{-0.2cm}
	{ \widetilde{\bf F}}= \frac{1}{\sqrt{N_t}}({\rm{cos}}{{\bf\Theta}}+j{\rm{sin}}{{\bf\Theta}}),
\end{equation}
where $j^2=-1$. 
Considering that the phase shifter is quantized, we quantize the output ${ \widetilde{\bf F}}$ in four bits after the network training.

Inspired by the PZF precoder \cite{bib2}, given the CSI $\bf H$ and analog precoding $\bf F$, the digital precoding is performed as 
\begin{equation}
	{\bf D} = {\bf H}_{\rm eq}({\bf H}_{\rm eq}{\bf H}_{\rm eq}^H)^{-1}{\bf \Lambda},
\end{equation}
where ${\bf H}_{\rm eq}\! \triangleq {\bf HF}$ is the equivalent channel and $\bf \Lambda$ is a diagonal matrix for column power normalization. The issue is that, from the DisNet, only implicit CSI $\widetilde{\bf H}$ is available, which means we are not safe to directly use the PZF algorithm or other traditional precoding calculations as in \cite{bib3}-\cite{bib7}. 
To address this issue, the last part of PreNet, namely BBNet as shown in Fig. 2, is designed to calculate the digital precoding from the output of RFNet, $\widetilde{\bf F}$, and the implicit CSI, $\widetilde{\bf H}$.

We notice that in the PZF algorithm, ${\bf H}_{\rm eq}$ is a diagonally dominant matrix when the channels of the UEs are nearly orthogonal.
Inspired by this feature, we define a non-linear function ${\rm{diag(\cdot)}}$ that extract all the $(i, i)$-th element of the ${\bf H}_{\rm{eq}}$, where $i = 1, ..., K$. 
If ${\bf H}_{\rm eq}$ is a diagonal matrix, then equation (11) is simplified to $ {\bf D} = {\rm diag}({\bf H}_{\rm eq}^{-1}){\bf \Lambda} = {\bf H}_{\rm eq}^{-1}{\bf \Lambda} $. 
Even if ${\bf H}_{\rm{eq}}$ is not a diagonally dominant matrix when the UEs are densely clustered, all the $(i,i)$-th elements of ${\bf H}_{\rm{eq}}$ are also the special elements of ${\bf H}_{\rm{eq}}$, because the physical meaning of analog-precoding lies in beam alignment. It may lead to a different distribution of ${\rm diag}({\bf H}_{\rm eq})$.
According to this conjecture, we divide BBNet into two smaller neural networks, i.e. \emph{Net a} and \emph{Net b} in Fig. 2. 
\emph{Net b} inputs ${\rm diag}({\bf H}_{\rm eq})$ and performs a rough estimate of the digital-precoding, while \emph{Net a} inputs ${\bf H}_{\rm eq}$ and calculates corrections for the digital-precoding.
In this way, the digital-precoding from BBNet, $\widetilde{\bf D}_u$, is calculated as
\begin{equation}
	\begin{split}
		\widetilde{\bf D}_u&={\bf D}_a + {\bf D}_b\\
		&={\sigma}({\bf W}_a\widetilde{{\bf H}}\widetilde{{\bf F}}+{\bf b})+{\sigma}({\bf W}_b{\rm{diag}}(\widetilde{{\bf H}}\widetilde{{\bf F}})+{\bf b}_b),
	\end{split}
\end{equation}
where $\sigma$ is the activation functions in (8), and ${\bf W}_a$, ${\bf b}_a$, ${\bf W}_b$, and ${\bf b}_b$ respectively represent the equivalent weights and equivalent biases of \emph{Net a} and \emph{Net b}. Before ending BBNet, we added a normalization module, which is defined as 
\begin{equation}
	\widetilde{\bf D} =\frac{\sqrt{K}\widetilde{{\bf D}}_u}{\|\widetilde{\bf F}\widetilde{{\bf D}}_u\|_{\rm F}}
\end{equation}
to meet the transmitting power constraint. 

\subsection{Network Training}
The proposed neural network realized the channel information compression which is a typical encoder-decoder problem. 
We cannot simply use supervised learning strategies like \cite{bib16}. Inspired by the work in \cite{bib8}, we propose a joint supervised-unsupervised learning approach. 
The entire training process is divided into two parts: pre-training and joint training.

During the pre-training stage, we build the pre-training network, DisNet and RFNet. DisNet and RFNet are trained sequentially by applying self-supervised learning and supervised learning. The loss function of DisNet is chosen as
\begin{equation}
	{ L}_1 = \frac{1}{KN_t}\|{\bf H}-\widetilde{\bf H}\|_{\rm F},
\end{equation}
and the loss funtion of RFNet is
\begin{equation}
	{ L}_2 = \frac{1}{KN_t}\|{\bf F}-\widetilde{\bf F}\|_{\rm F}.
\end{equation}
Note that when the DisNet is trained, the perfect CSI matrix $\bf H$ is fed into the DisNet to generate the imperfect CSI matrix $\widetilde{\bf H}$, as the input of RFNet. The training label $\bf F$ of RFNet is calculated from the perfect CSI using the previous method in \cite{bib2}. 

After the pre-training of DisNet and RFNet, we build the whole DNet, including DisNet, RFNet, and BBNet, and initialize the DisNet part and the RFNet part of this new DNet with the learned parameters. The BBNet of DNet is randomly initialized. We choose the sum rate of the MIMO system as the optimization objective of DNet. It follows
\begin{equation}
	{ L}_3=-\sum_{k=1}^{K}\log_2 \left (1+\frac{\frac{P}{K}|{\bf h}^H_k\widetilde{{\bf F}}\widetilde{{\bf d}}_k|^2}{1+\sum_{j\not=k}\frac{P}{K}|{\bf h}^H_k\widetilde{{\bf F}}\widetilde{{\bf d}}_j|^2}\right),
\end{equation}
where $\widetilde{{\bf d}}_i$ for $i=1,...,K$ is the $i$th column of $\widetilde{\bf D}$. During the joint training stage, we not only train the BBNet part but also modify the parameters of the DisNet and RFNet of the proposed DNet.
\subsection{Extension of DNet}
When the number of UEs changes, we can easily change the number of these encoder-decoder pairs without additional training, as for a given subcarrier, where the distribution of channels remains the same. 
Also, when the number of data streams is less than the number of RF chains, i.e., $K<N_{\rm{RF}}$, we suggest training the multiple groups of PreNet with different numbers of UEs in advance, then switching the network based on the actual number of UEs.
Meanwhile, although the system model and the corresponding problem formulation and solution are described for a narrowband system, which can directly extend to a wideband orthogonal frequency division multiplexing (OFDM) system, like in \cite{bibnew6}.

\section{Numerical Results}
\subsection{Dataset and Training}
We choose an open-source MIMO channel data set, called DeepMIMO \cite{bib17}, which is an OFDM system with 1024 subcarriers, as the source of training data set for making simulation experiments under more realistic channels. The number of UE is $4$ while the number of antennas at the BS is $64$. Considering that our model is based on a flat subband of an OFDM, which corresponds to a flat narrowband fading channel \cite{bib17}. Without loss of generality, we randomly sample the flat fading channel data on these subcarriers of the OFDM to form the training data set of our proposed DNet. 

\subsection{Achievable Sum Rates}
\begin{figure}[H]
\centering
\includegraphics[width = .7\textwidth]{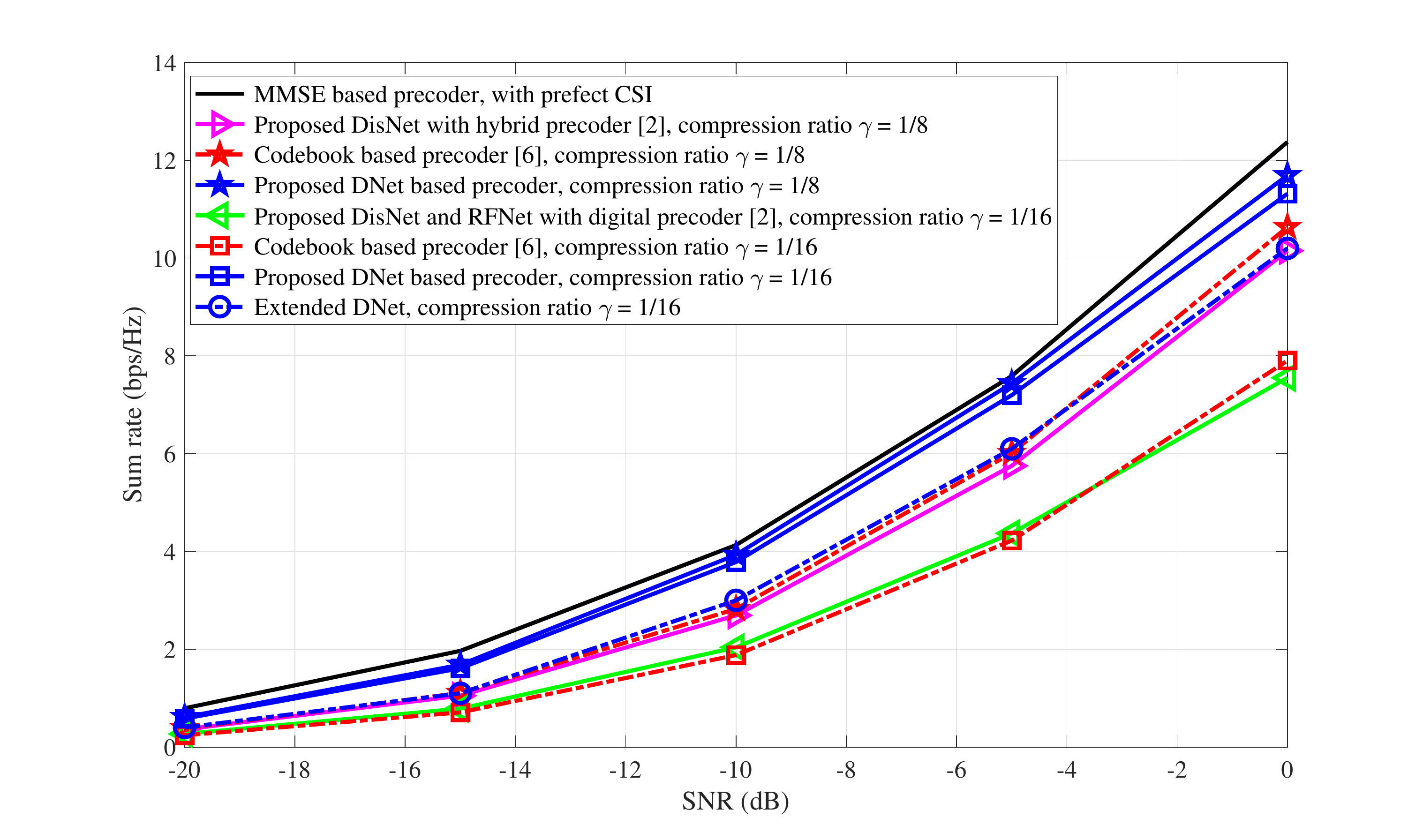}
\caption{Comparison of the achievable rate at different compression ratios.}
\label{fig:myphoto}
\end{figure}

In Fig. 3, we study the relationship between the sum rate and the SNR for different precoding schemes under various channel parameter compression ratios.
We see that the proposed scheme outperforms the other schemes, including the fully-digital MMSE precoding, the scheme using quantized codebook \cite{bib6}, the scheme combing the DisNet and hybrid precoding \cite{bib2}, the scheme combing the DisNet, RFNet, and digital precoding \cite{bib2}, and the scheme called extended DNet. This extended DNet first uses two out of four UEs’ channel data for the DisNet pre-training and then extends the pre-trained DisNet to the scenario with four UEs by sharing the same pre-trained parameters for the additional two UEs. The extended DNet shows the performance when more UEs are included.
For example, when compression ratio $\gamma = 1/16$ and the $\rm{SNR} = 0 \;dB$, the performance of the proposed algorithm is 3.1 bps/Hz higher than the traditional algorithm \cite{bib6}, and 3.5 bps/Hz higher than the scheme combing DisNet and hybrid precoding \cite{bib2}. The extended DNet achieves 2.1 bps/Hz higher sum rate than the conventional algorithm \cite{bib6}. This verifies our claim that in the case of limited feedback, the joint design of CSI compression and precoding improves the performance.

\begin{figure}[H]
	\centering
	\includegraphics[width = .7\textwidth]{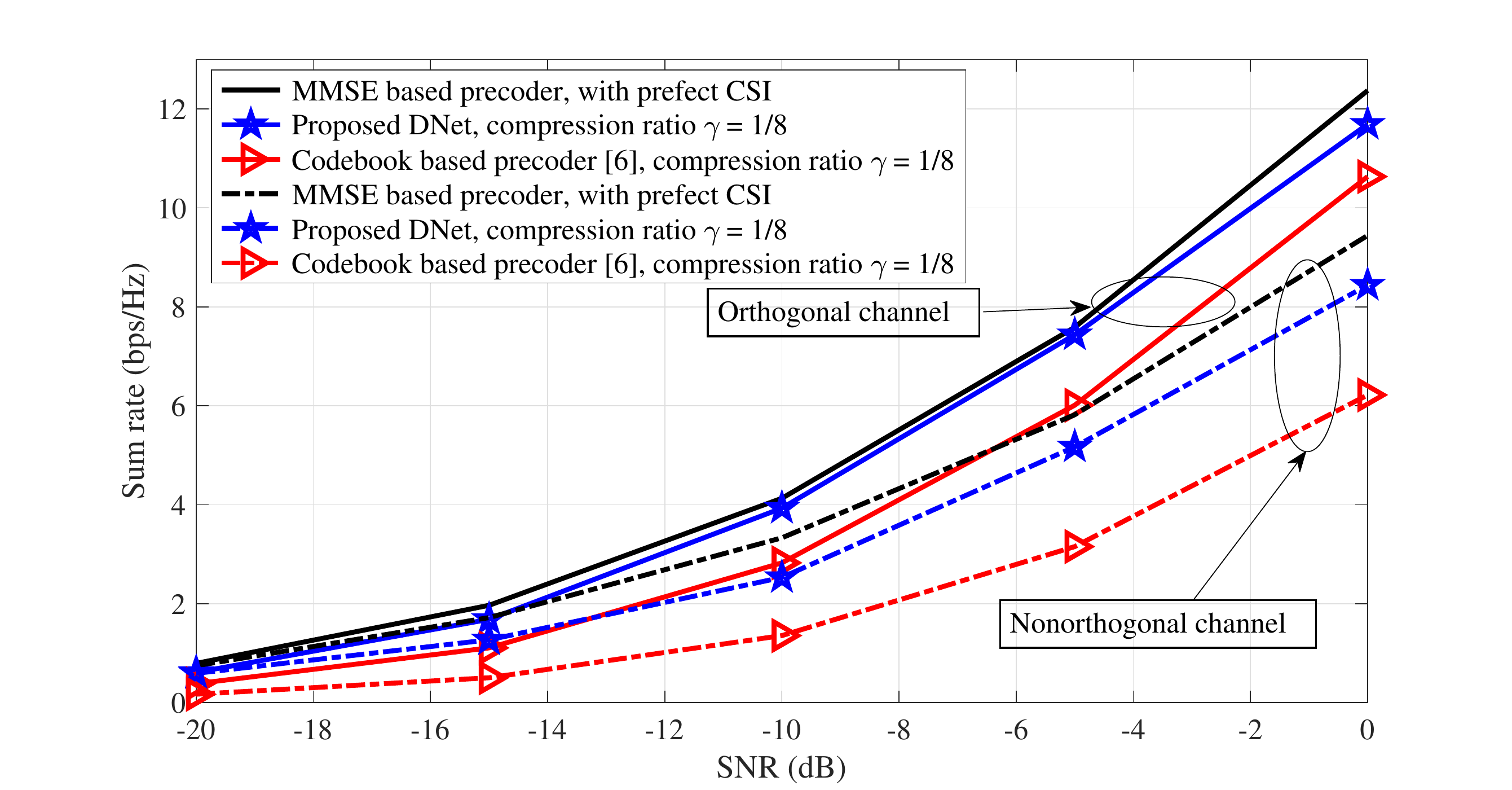}
	\caption{Comparison of the achievable rate under different channels.}
	\label{fig:myphoto}
\end{figure}

Then, we compare the performance of the hybrid precoding under orthogonal channels and nonorthogonal channels. As shown in Fig. 4, when $\gamma = 1/8$ and $\rm{SNR} = 0\; dB$, the performance degradation caused by the channel aggregation is more than 3 bps/Hz. However, when the channel is nonorthogonal, the achievable rate of the proposed DNet still outperforms the conventional algorithm in \cite{bib6} by 2.3 bps/Hz.

\section{Conclusion}

In this work, we proposed a distributed neural network called DNet and a corresponding training approach to calculate hybrid precoding with limited feedback in multiuser mmWave MIMO system. In this distributed network, all encoder-decoder network pairs are set to be the same, which helps reduce network complexity and improve network scalability. The joint design of the limited feedback network and the precoding network makes the performance of our proposed network significantly improved compared with the traditional codebook-based hybrid precoding algorithm.

\ifCLASSOPTIONcaptionsoff
  \newpage
\fi

\end{document}